\documentclass[12pt]{article}

\advance\voffset by -2.0cm \advance\hoffset by -1.25cm
\usepackage{amsmath}
\usepackage{amssymb}
\usepackage{amsthm}
\usepackage{amsfonts}
\theoremstyle{definition}

\newcommand{\CC}{\mathbb{C}} % Complessi
\newcommand{\RR}{\mathbb{R}} % Reali
\newcommand{\ZZ}{\mathbb{Z}} % Interi
\DeclareMathOperator{\tr}{tr} % traccia
\def\L{{\cal L}}

\newcommand{\II}{\mathbb{I}} % Matrice Unita'
\newcommand{\be}{\begin{equation}}
\newcommand{\ee}{\end{equation}}

\newlength{\oldcolsep}

\numberwithin{equation}{section}

\begin{document}

\title{An algorithm for the \\ Baker-Campbell-Hausdorff formula}
\author{Marco Matone}\date{}

\maketitle

\begin{center} Dipartimento di Fisica e Astronomia ``G. Galilei'' \\
 Istituto\\
Nazionale di Fisica Nucleare \\
Universit\`a di Padova, Via Marzolo, 8-35131 Padova,
Italy\end{center}

\begin{abstract}
\noindent A simple algorithm, which exploits the associativity of the BCH formula, and that can be generalized by iteration, extends the
remarkable simplification of the Baker-Campbell-Hausdorff (BCH) formula, recently derived by Van-Brunt and Visser. We show that if
 $[X,Y]=uX+vY+cI$, $[Y,Z]=wY+zZ+dI$, and, consistently with the Jacobi identity, $[X,Z]=mX+nY+pZ+eI$, then
$$
\exp(X)\exp(Y)\exp(Z)=\exp({aX+bY+cZ+dI})
$$
where $a$, $b$, $c$ and $d$ are solutions of four equations.
In particular, the Van-Brunt and Visser formula
$$\exp(X)\exp(Z)=\exp({aX+bZ+c[X,Z]+dI})
$$
extends to cases when $[X,Z]$ contains also elements different from $X$ and $Z$.
Such a closed form of the BCH formula may have interesting applications both in mathematics
and physics.
As an application, we provide the closed form of the BCH formula in the case of the exponentiation of the Virasoro algebra,
with  ${\rm SL}_2(\CC)$ following as a subcase.
We also determine three-dimensional subalgebras of the Virasoro algebra satisfying the Van-Brunt and Visser
condition. It turns out that the exponential form of ${\rm SL}_2(\CC)$ has a nice representation in terms of its eigenvalues and of the
 fixed points of the corresponding M\"obius transformation. This may have applications in Uniformization theory and Conformal Field Theories.

\end{abstract}

\newpage

\section{The algorithm}

\noindent Very recently Van-Brunt and Visser \cite{Van-Brunt:2015ala} (see also \cite{Van-Brunt:2015bza} for related issues) found
a remarkable relation that simplifies, in important cases, the Baker-Campbell-Hausdorff (BCH) formula. Namely, if $X$ and $Y$ are elements of a Lie
algebra  with commutator
\begin{equation}
[X,Y]=uX+vY+cI \ ,
\label{beoftheform}\end{equation}
with $I$ a central element and $u$, $v$, $c$, complex parameters, then \cite{Van-Brunt:2015ala}
\begin{equation}
\exp(X) \exp(Y)= \exp({X+Y+f(u,v)[X,Y]}) \ ,
\label{bbbvvv}\end{equation}
where
\be
f(u,v)=f(v,u)={(u-v)e^{u+v}-(ue^u-ve^v)\over uv(e^u-e^v)} \ .
\ee
In the following we formulate an algorithm which exploits the associativity of the BCH formula. Set $\alpha+\beta=1$ and consider the identity
\begin{equation}
\exp(X)\exp(Y) \exp(Z)= \exp(X)\exp({\alpha Y})\exp({\beta Y}) \exp(Z) \ .
\label{decomposizione}\end{equation}
If
\begin{equation}
[X,Y]=uX+vY+cI \ , \qquad [Y,Z]=wY+zZ+dI \ ,
\label{riecco}\end{equation}
then, associativity of the BCH formula and Eqs.(\ref{bbbvvv})(\ref{decomposizione}),  yield
\be
\exp(X) \exp(\alpha Y) = \exp({\tilde X}) \ , \qquad \exp(\beta Y)\exp(Z) = \exp({\tilde Y}) \ ,
\label{associativity}\ee
where
\begin{align}
\tilde X &:=g_{\alpha}(u,v)X+h_{\alpha}(u,v)Y+l_{\alpha}(u,v)cI \ , \cr
\tilde Y&:=h_{\beta}(z,w)Y+g_{\beta}(z,w)Z+l_{\beta}(z,w)dI \ ,
\label{colon1}\end{align}
with
$g_\alpha(u,v):=1+\alpha uf(\alpha u, v)$, $h_{\alpha}(u,v):=\alpha(1+vf(\alpha u, v))$ and $l_{\alpha}(u,v):=\alpha f(\alpha u, v)$.
Imposing
\begin{equation}
[\tilde X,\tilde Y]=\tilde u \tilde X+\tilde v \tilde Y+\tilde c I \ ,
\label{labella}\end{equation}
fixes $\alpha$, $\tilde u$, $\tilde v$ and $\tilde c$. This solves the BCH problem, since,
by (\ref{beoftheform}), (\ref{bbbvvv}), (\ref{decomposizione}) and (\ref{associativity})
\begin{equation}
\exp(X) \exp(Y) \exp(Z)= \exp(\tilde X) \exp(\tilde Y)=\exp({\tilde X+\tilde Y+f(\tilde u,\tilde v)[\tilde X,\tilde Y]}) \ .
\label{llasol}\end{equation}
Note that the commutator between $X$ and $Z$ may contain also $Y$
\begin{equation}
[X,Z]=m X+nY+pZ+e I \ .
\label{richiamare}\end{equation}
This is consistent with the Jacobi identity
\be
[X,[Y,Z]]+[Y,[Z,X]]+[Z,[X,Y]]=0 \ ,
\label{lajacobbbi}\ee
that
constrains $e, m,n$ and $p$ by a linear system.
Setting $Y=\lambda_0Q$ and $\lambda_-:=\lambda_0\alpha$, $\lambda_+:=\lambda_0\beta$, Eq.(\ref{decomposizione})
includes, as a particular case,
\begin{equation}
\exp(X) \exp(Z)=\lim_{\lambda_0\to 0} \exp(X)\exp({\lambda_- Q}) \exp({\lambda_+ Q}) \exp(Z) \ .
\label{decomposizionetre}\end{equation}
This explicitly shows  that the algorithm solves the BCH problem for $\exp(X) \exp(Z)$
in some of the cases when $[X,Z]$ includes elements of the algebra different from $X$ and $Z$.

\noindent The complete classification of the commutator algebras, leading to the closed form (\ref{llasol}) of the BCH formula, is investigated
in \cite{Matone:2015xaa}. The algorithm has been applied to the case of semisimple complex Lie algebras in \cite{Matone:2015oca}.

\noindent
In the next section we implement the above algorithm. In particular, we write down the linear system coming from the Jacobi identity and then find the
explicit expression for $\tilde c, \tilde u$ and $\tilde v$. We also find the equation, which is the key result of the algorithm, satisfied by $\alpha$.
In section 3, as an application, we consider the exponentiation
of the Virasoro algebra and derive the solution of the corresponding BCH problem. This includes, as a particular case, the
closed form of the BCH formula for ${\rm SL}_2(\CC)$.
We also determine three-dimensional subalgebras of the Virasoro algebra satisfying the Van-Brunt and Visser
condition (\ref{beoftheform}).  In the last section we apply the algorithm to find the exponential form of ${\rm SL}_2(\CC)$ matrices. Furthermore, we reproduce the same results, using an alternative method,
and extending them to ${\rm GL}_2(\CC)$ matrices. In this respect, it seems that in the literature, similar expressions for $\gamma\in{\rm SL}_2(\CC)$ are usually given only separately for three distinguished cases, depending if $\gamma_{11}^2+\gamma_{12}\gamma_{21}$
is negative, vanishing or positive. It turns out that the expression of $X$ in $\gamma=\exp(X)$ has a geometrical representation which is not directly evident in $\gamma$. Namely, it turns out that $X$ can be nicely expressed
in terms of its eigenvalues and of the fixed points $z_\pm$, solutions of the equation $z=\gamma z$, where $\gamma z$ is the M\"obius transformation
\be
\gamma z:={\gamma_{11}z+\gamma_{12}\over \gamma_{21}z+\gamma_{22}} \ .
\ee
Such a geometrical representation of $\gamma$ may be of interest in the framework of Uniformization theory and Conformal Field Theories.

\section{Implementation of the algorithm}

Let us write down the linear system that follows by the Jacobi identity (\ref{lajacobbbi})
\begin{align}
uw+mz & =0 \ ,  \cr
vm-wp + n(z-u) & = 0 \ ,  \cr
pu+zv & = 0 \ ,  \cr
c(w+m)+e(z-u)-d(p+v)&=0 \ .
 \label{system}\end{align}

\noindent
Replacing $\tilde X$ and $\tilde Y$ on the right hand side of (\ref{labella}) by their expressions in terms of $X$, $Y$ and $I$, and comparing the result
with the direct computation, by (\ref{riecco}), (\ref{colon1}) and (\ref{richiamare}), of $[\tilde X,\tilde Y]$, yields
\begin{align}
\tilde c&=(h_\beta (z,w)-g_\beta(z,w)l_\alpha(u,v) m) c +(h_\alpha(u,v)-g_\alpha(u,v) l_\beta(z,w) p)d+g_\alpha(u,v)g_\beta(z,w) e \ ,  \cr
\tilde u&=h_{\beta}(z,w)u+g_{\beta}(z,w)m \ ,   \cr
\tilde v&=g_{\alpha}(u,v)p+h_{\alpha}(u,v)z \ , \cr
\tilde u & h_{\alpha}(u,v)+\tilde vh_{\beta}(z,w)=g_{\alpha}(u,v)h_{\beta}(z,w)v+g_{\alpha}(u,v)g_{\beta}(z,w)n +h_{\alpha}(u,v)g_{\beta}(z,w)w \ .
\label{perlaconsistenza}\end{align}
The first three equations fix $\tilde c$, $\tilde u$ and $\tilde v$ in terms of $\alpha=1-\beta$.
Replacing the expressions of $\tilde u$ and $\tilde v$ in the fourth equation
provides the following equation for $\alpha$
\begin{equation}
h_{\alpha}(u,v) [h_{\beta}(z,w)(u+z)+g_{\beta}(z,w)(m-w)]+g_{\alpha}(u,v)
[h_{\beta}(z,w)(p-v)-g_{\beta}(z,w)n]=0 \ .
\label{fonda}\end{equation}
This is the basic equation of the algorithm and
is further investigated, together with the linear system (\ref{system}), in
\cite{Matone:2015xaa}.
Note that
\begin{align}
g_\alpha(u,v)={v-\alpha u\over v}{e^{\alpha u/2}\sinh(v/2)\over \sinh[(v-\alpha u)/2] } \ ,  \cr
h_{\alpha}(u,v)={v-\alpha u\over u}{e^{v/2}\sinh(\alpha u/2)\over \sinh[(v-\alpha u)/2] } \  .
\end{align}
We then obtained a closed form of the BCH formula for cases in which the commutator contains other elements than the ones in the commutator.
The above algorithm can be extended to more general cases, e.g. by considering
decompositions like the one in (\ref{decomposizione}) for
\be
\exp({X_1}) \cdots \exp({X_n}) \ .
\ee

\section{Exponentiating the Virasoro algebra}

In this section we apply the algorithm, leading to closed forms of the BCH formula, in the case of the exponentiation
of the Virasoro algebra
\be
[ {\L}_j, {\L}_k]=(k-j){\L}_{j+k}+{c\over 12}(k^3-k)\delta_{j+k,0}I \ ,
\label{lavv}\ee
$j,k\in\ZZ$. In particular, we find the closed form for $W$ in
\be
\exp(X) \exp(Y) \exp(Z)=\exp(W) \ ,
\ee
where
\be
X:=\lambda_{-k} {\L}_{-k} \ , \qquad Y:=\lambda_0 {\L}_0 \ , \qquad Z:=\lambda_k {\L}_k \ .
\label{kappas}\ee
This is particularly interesting because we do not know alternative ways to get it.
The case of ${\rm SL}_2(\CC)$ follows straightforwardly since the ${\rm sl}_2(\RR)$  algebra
\begin{equation}
[L_j,L_k]=(k-j)L_{j+k} \ ,
\end{equation}
$j,k=-1,0,1$,  is a subalgebra of (\ref{lavv}).
Note that setting
$E_-=L_1$, $H=-2L_0$ and $E_+=-L_{-1}$,
reproduces the other standard representation of the ${\rm sl}_2(\RR)$ algebra $[H,E_+]=2E_+$, $[E_+,E_-]=H$ and $[H,E_-]=-2E_-$.
Note that,
\be
[X,Y]=k\lambda_0 X \ , \qquad [Y,Z]=k\lambda_0 Z \ , \qquad  [X,Z]=\lambda_{-k}\lambda_k\Big[{2k\over \lambda_0} Y+
{c\over 12} (k^3-k)\Big]  \ ,
\ee
where, besides $c=d=v=w=0$, we have, consistently with the Jacobi identity, $m=p=0$. The other commutator parameters are
\begin{align}
u & =z=k\lambda_0 \ , \cr
n & =\lambda_{-k}\lambda_k{2k\over\lambda_0} \ , \cr
e & =\lambda_{-k}\lambda_k{c\over 12}(k^3-k) \ .
\label{centrrr}\end{align}
By (\ref{fonda}) it follows that $\alpha$ satisfies the equation
\be
ng_\alpha(u,0)g_\beta(u,0)=2uh_\alpha(u,0)h_\beta(u,0) \ ,
\ee
so that, using
\be
h_\alpha(u,0)=\alpha \ , \qquad g_\alpha(u,0)={\alpha u\over 1-e^{-\alpha u}} \ ,
\ee
and recalling that $\lambda_-:=\lambda_0\alpha$, $\lambda_+:=\lambda_0\beta$, one gets
\be
{}e^{-k\lambda_{\pm}}=
{1+e^{-k\lambda_0}-k^2\lambda_{-k}
 \lambda_k\pm \sqrt{(1+e^{-k\lambda_0}-k^2\lambda_{-k}\lambda_k)^2-4e^{-k\lambda_0}}\over2} \ .
\label{eallakappa}\ee
Next, observe that, by (\ref{perlaconsistenza}), $\tilde u=k\lambda_+$, $\tilde v=k\lambda_-$, and
$c_k\equiv\tilde c=eg_\alpha(k\lambda_0,0)g_{\beta}(k\lambda_0,0)$, that is
\be
c_k={\lambda_-\lambda_{-k}\over 1-e^{-k\lambda_-}}{\lambda_+\lambda_k\over 1-e^{-k\lambda_+}}{c\over 12}(k^5-k^3) \ ,
\label{lattllc}\ee
that, by (\ref{eallakappa}), is equivalent to
\be
{}c_k={\lambda_{-k}\lambda_k\over \lambda_+-\lambda_-}
  \bigg({\lambda_+\over 1-e^{-k\lambda_+}}-{\lambda_-\over 1-e^{-k\lambda_-}}\bigg){c\over12}(k^4-k^2) \ .
\ee
Finally, Eq.(\ref{llasol}) yields
$$  \exp({\lambda_{-k}{\L}_{-k}})\exp({\lambda_0{\L}_0})
\exp({\lambda_k{\L}_k})= $$
\begin{align}
&\exp\Big\{{\lambda_+-\lambda_-\over e^{-k\lambda_-}-e^{-k\lambda_+}}\Big[k\lambda_{-k}{\L}_{-k}+\Big(2-e^{-k\lambda_+}-e^{-k\lambda_-}\Big){\L}_0+k\lambda_k {\L}_k+c_kI\Big]\Big\} \ .
\label{viralasoluzionee}\end{align}
In the case $\lambda_0=0$ we have
\be
\exp({\lambda_{-k}{\L}_{-k}})
\exp({\lambda_k{\L}_k})=\exp\bigg[ {\lambda_+\over \sinh (k\lambda_+)}(k\lambda_{-k}{\L}_{-k}+k^2\lambda_{-k}\lambda_k{\L}_0+k\lambda_k {\L}_k+c_kI)\bigg]\ ,
\label{viralasoluzioneezeroBBB}\ee
with
\begin{equation}
c_k=\lambda_{-k}\lambda_k{c\over24}(k^4-k^2) \ .
\end{equation}
Let us report the relevant case corresponding to ${\rm SL}_2(\CC)$, obtained by setting $k=1$ in the above formulas. We have
$$  \exp({\lambda_{-1}{L}_{-1}})\exp({\lambda_0{L}_0})
\exp({\lambda_1{L}_1})= $$
\begin{align}
&\exp\Big\{{\lambda_+-\lambda_-\over e^{-\lambda_-}-e^{-\lambda_+}}\Big[\lambda_{-1}{L}_{-1}+\Big(2-e^{-\lambda_+}-e^{-\lambda_-}\Big){L}_0+\lambda_1 {L}_1\Big]\Big\} \ ,
\label{viralasoluzioneeAAA}\end{align}
and
\be
\exp({\lambda_{-1}{L}_{-1}})
\exp({\lambda_1{L}_1})=\exp\bigg[ {\lambda_+\over \sinh (\lambda_+)}(\lambda_{-1}{L}_{-1}+\lambda_{-1}\lambda_1{L}_0+\lambda_1 {L}_1)\bigg]\ .
\label{viralasoluzioneezeroBBBAAA}\ee

\noindent
Interestingly, the algorithm for the BCH formula may be extended to any three-dimensional subalgebras of the Virasoro algebra. In this respect, it can be easily seen that
the highest finite dimensional subalgebras of the Virasoro are the four-dimensional ones generated by $\L_{-n}$, $\L_0$, $\L_n$ and the central element $I$, for all $n\in \ZZ\backslash\{0\}$.
Above we solved the BCH problem for all such subalgebras. Since all the remanent non-trivial Virasoro subalgebras are either three-dimensional, each one containing $I$, or two-dimensional,
it follows that all of them
satisfy the condition (\ref{beoftheform}) and therefore the corresponding BCH problem of finding $Z$ such that $\exp(X)\exp(Y)=\exp(Z)$ is easily solved by (\ref{bbbvvv}).
In this respect, note that it can be easily seen that the two dimensional
subalgebras are generated by $\L_n$ and $\L_0$, for all $n\in\ZZ\backslash\{0\}$.
A two-parameter family of three-dimensional subalgebras of the Virasoro algebra, is the one where each subalgebra is
generated by
\begin{equation}
 X_n(\delta,\epsilon):=\delta{\L}_0+\epsilon {\L}_n \ , \qquad X_{-n}(\epsilon,\delta) \ , \qquad I  \ ,
\end{equation}
whose commutator is
\begin{equation}
[X_n(\delta,\epsilon),X_{-n}(\epsilon,\delta)]=-n\epsilon X_n(\delta,\epsilon) -n{\delta} X_{-n}(\epsilon,\delta) +\delta\epsilon {c\over 12}(n^3-n)I\ .
\end{equation}
Other three-dimensional subalgebras of the Virasoro algebra are the one-parameter family of subalgebras, each one generated by
\begin{equation}
 X_n(\alpha):={\L}_{2n}+\alpha {\L}_n+{2\over 9}{\alpha^2}{\L}_0 \ , \qquad Y_{-n}(\alpha):={\L}_{-n}+{3\over\alpha}{\L}_0 \ , \qquad I \ ,
\end{equation}
whose commutator is
\begin{equation}
[X_n(\alpha),Y_{-n}(\alpha)]=-6{n\over \alpha}X_n(\alpha)-{2\over9}{n\alpha^2} Y_{-n}(\alpha)+\alpha{c\over12}(n^3-n)I \ .
\end{equation}

\section{Geometrical constructions}

Note that, in the case of ${\rm SL}_2(\RR)$, therefore for real $\lambda_k$'s, depending on the values of the $\lambda_k$'s,
the factor ${\lambda_+-\lambda_-\over e^{-\lambda_-}-e^{-\lambda_+}}$ in (\ref{viralasoluzioneeAAA}) may take complex values.
This is in agreement with the non-surjectivity of the exponential map for ${\rm sl}_2(\RR)$ into ${\rm SL}_2(\RR)$.
In particular, exponentiating ${\rm sl}_2(\RR)$ cannot give  ${\rm SL}_2(\RR)$ matrices whose trace is less than $-2$.
The case with trace $-2$ and non-diagonalizable matrices is critical, both for ${\rm SL}_2(\RR)$ and ${\rm SL}_2(\CC)$.

\noindent Let us express Eq.(\ref{viralasoluzioneeAAA}) in terms of the associated ${\rm SL}_2(\CC)$ matrix
\begin{equation}
\gamma=\left(\begin{array}{c}A\\
C\end{array}\begin{array}{cc}B\\ D
\end{array}\right) \ .
\end{equation}
Replace then the $L_k$'s in the left hand side of (\ref{viralasoluzioneeAAA}) by their matrix representation
\begin{equation}L_{-1}=\left(\begin{array}{c}0\\
0\end{array}\begin{array}{cc}-1\\ 0
\end{array}\right) \ , \qquad
L_{0}=\left(\begin{array}{c}-{1\over2}\\
0\end{array}\begin{array}{cc}0\\ {1\over2}
\end{array}\right) \ , \qquad L_{1}=\left(\begin{array}{c}0\\
1\end{array}\begin{array}{cc}0\\ 0
\end{array}\right) \ ,
\end{equation}
using
\begin{equation}\exp({L_{-1}})=\left(\begin{array}{c}1\\
0\end{array}\begin{array}{cc}-1\\ 1
\end{array}\right) \ , \quad
\exp({L_{0}})=\left(\begin{array}{c}e^{-{1\over2}}\\
0\end{array}\begin{array}{cc}0\\ e^{1\over2}
\end{array}\right) \ , \quad \exp({L_{1}})=\left(\begin{array}{c}1\\
1\end{array}\begin{array}{cc}0\\ 1
\end{array}\right) \ .
\end{equation}
Comparing the result with $\gamma$ yields
\begin{equation}
A=e^{\lambda_0/2}(e^{-\lambda_0}-\lambda_{-1}\lambda_1) \ ,  \qquad B=-\lambda_{-1}e^{\lambda_0/2} \ ,  \qquad C=\lambda_1e^{\lambda_0/2} \ ,  \qquad D=e^{\lambda_0/2} \ ,
\end{equation}
so that
\begin{equation}
e^{-\lambda_{\pm}}={t\pm \sqrt{t^2-1}\over D} \ ,
\label{incredible}\end{equation}
where $t:={1\over2}\tr \gamma$.
Since the eigenvalues of $\gamma$,  solutions of the characteristic polynomial
$\nu^2-2t\nu+1=0$, are
$\nu_\pm={t\pm \sqrt{t^2-1}}$,
we have
\begin{equation}
e^{-\lambda_{\pm}}={\nu_\pm\over D} \ ,
\end{equation}
so that
\begin{equation}
\exp(\lambda_{-1}L_{-1})\exp(\lambda_0L_0)
\exp(\lambda_1L_1)=\exp\bigg\{{\ln (\nu_+/\nu_-)\over \nu_+-\nu_-}[CL_{1}+(D-A)L_0-BL_{-1}]\bigg\} \ .
\label{formulaff}\end{equation}
Note that when $\gamma$ is parabolic, that is for $|t|=1$, and therefore $\nu_+=\nu_-$, we have
\begin{equation}
{\ln (\nu_+/\nu_-)\over \nu_+-\nu_-}=2\,{\rm sgn}(t) \ .
\end{equation}
Also, note that when $\gamma$ is elliptic, that is for $|t|<1$, one has $\nu_+=\bar\nu_-=\rho e^{i\theta}$,
so that
\begin{equation}{\ln (\nu_+/\nu_-)\over \nu_+-\nu_-}={\theta\over \rho\sin \theta} \ .
\end{equation}
Eq.(\ref{formulaff}) implies
\begin{equation}\left(\begin{array}{c}A\\
C\end{array}\begin{array}{cc}B\\ D
\end{array}\right)=\exp\bigg[{{\ln (\nu_+/\nu_-)\over \nu_+-\nu_-}\left(\begin{array}{c}(A-D)/2\\
C\end{array}\begin{array}{cc}B\\ (D-A)/2
\end{array}\right)}\bigg] \ ,
\label{special}\end{equation}
equivalently, using $\nu_+\nu_-=1$,
\begin{equation}
\gamma= \exp\bigg[{\ln (t+\sqrt{t^2-1})\over \sqrt{t^2-1}}(\gamma-tI_2)\bigg] \ ,
\label{formulao}\end{equation}
which is indefinitely iterable by replacing $\gamma$ on
the right hand side, by its exponential form, that is by the expression on the right hand side itself.

\noindent The relation (\ref{formulao}) can be derived in an alternative way. First note that $\gamma=\exp(X)$ does not uniquely fix $X$. For example,
$\exp(X)=\exp(X)\exp(2\pi i k I_2)=\exp(X+2\pi i kI_2)$, $k\in \ZZ$. However, $X$ can be consistently fixed to be traceless, so that, being $X^2$ proportional to $I_2$, one has
$\gamma = \exp(X)=aX+tI_2$. Therefore $\gamma=\exp[a^{-1}(\gamma-tI_2)]$ for some $a$. For distinct eigenvalues the diagonalization of both sides (by the same
matrix), reproduces (\ref{formulao}), since it fixes
\begin{equation}
a={\nu_+-\nu_-\over 2\ln\nu_+} \ .
\end{equation}
A particular case of (\ref{formulao}) is when $D=A^{-1}$
\begin{equation}
\gamma=\exp\bigg[{\ln A\left(\begin{array}{c}1\\
{2C\over A-A^{-1}} \end{array}\begin{array}{cc} {2B\over A-A^{-1}}\\ -1
\end{array}\right)}\bigg] \ ,
\end{equation}
where, since $AD-BC=1$, either $B=0$ or $C=0$. It follows, that when $D=A^{-1}$, with $A^2=1$,  $\gamma$ admits exponentiation  only if $A=1$.
For example, $\left(\begin{array}{c}-1\\
0\end{array}\begin{array}{cc}1\\ -1
\end{array}\right)$ cannot be expressed as the exponential neither of ${\rm sl}_2(\RR)$ nor of ${\rm sl}_2(\CC)$. Of course, this is not a problem in the case such a matrix
is seen as an element of ${\rm PSL}_2(\RR)={\rm
SL}_2(\RR)/\{\pm\II\}$.

\noindent Let us derive a more geometrical representation of (\ref{formulao}), useful, e.g., in the framework of Uniformization Theory and in Conformal Field Theories.
Consider the M\"obius transformation
\begin{equation}
\gamma z:= {Az+B\over Cz+D} \ ,
\end{equation}
and the solutions of the fixed point equation $\gamma z =z$
\begin{equation}
z_\pm= {(A-D)/2\pm\sqrt{t^2-1}\over C} \ .
\end{equation}
If $B=0$, then
\begin{equation}\left(\begin{array}{c}A\\
C\end{array}\begin{array}{cc}0\\ D
\end{array}\right)=\exp\bigg[{{\ln (\nu_+/\nu_-)\over z_+-z_-}\left(\begin{array}{c}z_++z_-\\
1\end{array}\begin{array}{cc}0\\ -z_+-z_-
\end{array}\right)}\bigg] \ ,
\end{equation}
otherwise, in the case of  M\"obius transformations, one can fix $B=1$ to get
\begin{equation}\left(\begin{array}{c}A\\
C\end{array}\begin{array}{cc}1\\ D
\end{array}\right)=\exp\bigg[{{\ln (\nu_+/\nu_-)\over z_+-z_-}\left(\begin{array}{c}z_++z_-\\
1\end{array}\begin{array}{cc}{z_+-z_-\over\nu_+-\nu_-}\\ -z_+-z_-
\end{array}\right)}\bigg] \ .
\end{equation}
We conclude by observing that our findings trivially extend to ${\rm GL}_2(\CC)$. In particular, multiplying both sides of (\ref{formulao}) by $\sqrt{|\gamma|}I_2$, $|\gamma|:=\det\gamma$, one gets for $\gamma\in{\rm GL}_2(\CC)$
\begin{equation}
\gamma= \exp\bigg[{\ln (t+\sqrt{t^2-|\gamma|})\over \sqrt{t^2-|\gamma|}}(\gamma-tI_2)+{1\over2}\ln(|\gamma|)I_2\bigg] \ .
\label{formulaoGL}\end{equation}

\section*{Acknowledgements} It is a pleasure to thank  Pieralberto Marchetti,  Leonardo Pagani, Paolo Pasti,
 Dmitri Sorokin and Roberto Volpato for interesting
discussions.


\newpage

\begin{thebibliography}{99}


%\cite{Van-Brunt:2015ala}
\bibitem{Van-Brunt:2015ala}
  A.~Van-Brunt and M.~Visser,
  %``Special-case closed form of the Baker-Campbell-Hausdorff formula,''
  arXiv:1501.02506. % [math-ph].
  %%CITATION = ARXIV:1501.02506;%%
  %1 citations counted in INSPIRE as of 17 Feb 2015

%\cite{Van-Brunt:2015bza}
\bibitem{Van-Brunt:2015bza}
  A.~Van-Brunt and M.~Visser,
  %``Simplifying the Reinsch algorithm for the Baker-Campbell-Hausdorff series,''
  arXiv:1501.05034. %[math-ph].
  %%CITATION = ARXIV:1501.05034;%%

%\cite{Matone:2015xaa}
\bibitem{Matone:2015xaa}
  M.~Matone,
  %``Classification of Commutator Algebras Leading to the New Type of Closed Baker-Campbell-Hausdorff Formulas,''
  arXiv:1503.08198. %[math-ph].
  %%CITATION = ARXIV:1503.08198;%%

%\cite{Matone:2015oca}
\bibitem{Matone:2015oca}
  M.~Matone,
  %``Closed Form of the Baker-Campbell-Hausdorff Formula for Semisimple Complex Lie Algebras,''
  arXiv:1504.05174. %[math-ph].
  %%CITATION = ARXIV:1504.05174;%%




\end{thebibliography}
\end{document}